\documentclass{svmult}
\usepackage{epsfig,graphicx} 
\usepackage[bottom]{footmisc}
\usepackage{natbib,url}      
\usepackage{bm}
\usepackage{makeidx}
\bibpunct{(}{)}{;}{a}{}{,}   
\def\cite#1{\citealp{#1}}    
\def\authorindex#1{}
\usepackage{ifpdf}
\begin{document}\newcount\preprintheader\preprintheader=1
\title*{High-resolution spectroscopy of  the   R Coronae Borealis 
and Other Hydrogen Deficient Stars}
\titlerunning{High-resolution spectroscopy of  the   R Coronae}

\author{N. Kameswara Rao\inst{1} 
        \and  
        David L. Lambert\inst{2}}
\institute{Indian Institute of Astrophysics, Bangalore 560034, India.\\
          \texttt{nkrao@iiap.res.in}
          \and
         The W.J. McDonald Observatory, The University of Texas, Austin, 
         TX 78712-1083, USA \\ \texttt{dll@astro.as.utexas.edu}}
\maketitle

\begin{abstract}

  High-resolution  spectroscopy is a very important tool for studying
stellar physics, perhaps, particularly so for such enigmatic objects like  the
R Coronae Borealis and related Hydrogen deficient stars that produce
carbon dust in addition to their peculiar abundances. 
 Examples of how high-resolution spectroscopy is used in the study of
these stars to address the two major puzzles are presented: (i) How are
such rare H-deficient stars created? and (ii) How and where are the
obscuring soot clouds produced around the R Coronae Borealis stars?
\end{abstract}

\section{Introduction}
We  congratulate the organizers for arranging and conducting this
conference, celebrating the 150 years of the discovery by (Gustav Robert)
Kirchoff and (Robert Wilhelm Eberhard) Bunsen that various gases can be 
easily and positively identified by a detailed study of the light they emit 
and absorb.  As Paul Merrill (\cite{m0}) noted, a new era in astronomy
began when Bunsen saw `in yellow flame of an ordinary alcohol lamp whose wick
was sprinkled with salt, and possibilities of the chemical analysis of the most 
distant stars'. Thus began the era of astronomical spectroscopy. 
Merrill's own contributions to our discipline are legendary with his
discovery of Tc\,{\sc i} lines in the spectra of S stars signaling that
element synthesis in stars was a continuing phenomena (\cite{m1}).
  
A few historical remarks might be appropriate.  Astronomical spectroscopy in 
India started during the famous total solar eclipse that occurred on 1868 
August 18 in which a spectroscope was first used to study the nature of solar 
prominences leading to the
discovery of helium.  Madras observatory, from which the present host
Institution evolved, also played a role. Norman Robert Pogson, then director
of the observatory, used a visual spectroscope to look at the prominences 
and noted a bright line in the yellow close to but not
coincident with sodium D lines.  This was the D3 line - the same line
as found by Janssen during the 1868 eclipse but by Lockyer without the 
aid of eclipse.  The first stellar spectrum  recorded in India was 
obtained at Madras (now Chennai); this was a spectrum
of the Wolf-Rayet star $\gamma^2$ Velorum observed in 1871.

Spectroscopic  analysis of some of the most enigmatic stars yet discovered
is the theme of our contribution. The chosen stars are the R Coronae Borealis
stars and their putative cooler relatives, the hydrogen-deficient cool
carbon stars (HdCs). The enigma presented by these stars has two principal
parts: (i) How did these stars become so H-deficient? and (ii) how do the
RCBs but not the HdCs develop thick clouds of soot that obscure the
stellar photosphere from view? Answers to these questions are emerging.
Here, we highlight particular contributions to their solution from
analysis of high-resolution optical and infrared spectra. 
\citet{Disney2000} has lamented (tongue in cheek, we suppose) that `the 
tragedy of Astronomy is that most information lies in spectra'.
We would counter by substituting `allure' for `tragedy' and add that
one should expect to apply all available observational tools in seeking
answers to our questions.

R Coronae Borealis stars are a rare class of peculiar variable stars. The two 
defining characteristics of RCBs are (i) a propensity to fade at unpredictable 
times by up to about 8 magnitudes as a result of obscuration by clouds of soot,
and (ii) a supergiant-like atmosphere that is very H-deficient and He-rich. 
The $T_{\rm eff}$ of the class ranges from 8000 to 4000 K and $\log g$ from 0.5 to 1.5 
(cgs units) except for the two hotter stars DY Cen (19000 K) and MV Sgr 
(14000 K).  The M$_{\rm V}$ ranges from -5 to -2.5 and $\log L$ from  4.0 to 3.2 
L$_{\odot}$. Presently, there are 52 known RCBs in the Galaxy 
(\cite{nkr5,ca2009}), 21 known members in
the Large Magellanic Cloud and 6 in the Small Magellanic Cloud. 
The total number of RCBs in the Galaxy may be about 3200 (\cite{nkr1})
There are only 5 known HdC stars
in the Galaxy. Their detection  may be hampered because they do not show 
RCB-like light variations, an infrared excess or a  stellar wind
(\cite{nkr4}). They overlap the  RCBs in T$_{\rm eff}$ with a spread from
about  6500 to 5000 K. At the high $T_{\rm eff}$ end of the RCBs sit the
Extreme helium stars (EHes) of which 21 are known in the
Galaxy with $T_{\rm eff}$
from 32000 to 9000 K and $\log g$ from 0.75 to 4.0. Their $\log L$ ranges from
4.4 to 3.0 L$_{\odot}$. 
Here, we give no more than passing attention to the EHes and other, even
hotter H-deficient stars that may form an evolutionary
sequence with the HdCs and RCBs.

A fascinating aspect of RCBs is that they present several faces  to observers,
like the Hindu mythological god Karthikeya (son of Kritthikas -Pleiades) 
with his six faces. 
The faces presented to our view are (i) a chemically peculiar supergiant,
(ii) a variable star  obscured without warning from our view by clouds of
soot, (iii) a small amplitude
pulsation both in light and radial velocity with a period around 40 days, 
(iv) an infrared 
source producing dust and blowing it away after several days, weeks, months 
or even years as clouds,
(v) a  hot stellar wind, and (vi) and may even be a central star of a low density
nebula as evidenced by the presence of nebular lines of [O II], [N II] (Fig. 1)
and [S II]. They also have  putative relatives of higher and lower
temperatures, as intriguing as Karthikeya's relatives.

Our emphasis is on insights obtainable from high-resolution spectra with the
assumption that relevant atomic and molecular data of adequate precision and
completeness are to be provided from laboratory and theoretical spectroscopic
studies. This assumption is patently false, as is no surprise to
practicing astronomical spectroscopists. On the chance that a reader of this
paper may wish to take up a challenge, we mention two aspects of the
incompleteness of knowledge of the spectrum of neutral carbon that
compromise some analyses of the spectra of RCBs. 

Spectra of RCBs are crossed by many lines of C\,{\sc i}, as
noted by \citet{keenan}. One might refer to the
spectrum of a RCB as arising from a column of somewhat contaminated
high-temperature carbon vapour. A great step forward in identifying the
C\,{\sc i} lines was made when \citet{Johansson} extended the laboratory
spectrum and knowledge of the C\,{\sc i} term system. With modern
spectra of RCBs, many suspected C\,{\sc i} lines remain unidentified and cry out
for a new laboratory investigation of the C\,{\sc i} spectrum.
Today, astronomical spectroscopists often require more than a set of
wavelengths and a table of energy levels. This is certainly true for
the pursuit of an abundance analysis leading to the chemical
composition of a RCB. Abundance analysis at a minimum requires 
knowledge of the $gf$-values for identified lines including the
C\,{\sc i} lines. \citet{nkr2} report on an extensive
abundance analysis of RCBs and their uncovering of a `carbon problem'
which may arise because the adopted theoretical $gf$-values 
or the photoionization cross-sections for neutral carbon  were too
large by a factor of about four. (Asplund et al. also suggested
possible ways to reduce the carbon problem without appealing to
deficiencies in atomic physics.) Where are the physicists who will
attack the spectrum of the carbon atom with the precision and
thoroughness that we seek?

\begin{figure}
\center
\includegraphics[height=7cm,width=9cm]{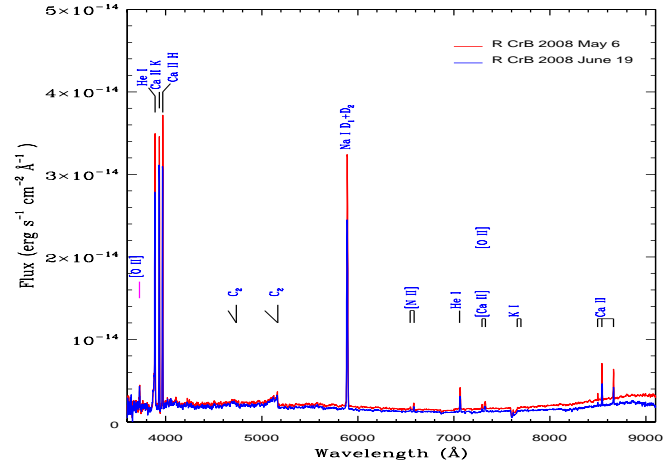}
\caption{Low resolution spectra obtained on 2008 May 6 and 2008 June 19
 with the Himalayan Chandra Telescope (HCT)
 during the  current  minimum of R CrB which began in 2007. 
Emission lines are identified. A variety of emission lines both broad and sharp 
are present during deep minimum. Several forbidden lines of [O II], [N II] 
 representing low density nebular gas are also present (see text).}
\end{figure}

\section{Major Puzzles}

As stated in the Introduction, the two principal puzzles presented by the RCBs
concern their origins and their ability to produce soot clouds. The HdCs
share the puzzle about their origins but not that of soot production.           
Of course, the `dust' puzzle might incorporate the HdCs with the RCBs by
asking `Why is it that RCBs but not the HdCs, which are on average cooler than the
RCBs, are prone to soot production?'

\subsection{Origins}

The basic problem is to understand how a H-deficient star evolves
from a H-rich single or binary star; 
H-poor He-rich gas clouds capable of sustaining stellar nurseries
have not been found.
Complete resolution of the problem would show how the classes of
H-deficient stars -- HdCs - RCB - EHe - beyond -- are or are not
related.

Presently, there are two principal hypotheses for the origins:(i) the
double-degenerate (DD) scenario and (ii) a late thermal pulse or final
flash (the FF scenario) in a luminous star on the white-dwarf cooling track.

In the DD scenario,  a He white dwarf
is accreted by a C-O white dwarf as loss of orbital energy drives the
white dwarfs to merge. There  are rather specific ranges of initial masses and
separations for normal (H-rich) stars in a binary to merge and
produce a H-deficient He-rich star. (Other combinations produce other
odd stars such as a single white dwarf with a mass exceeding Chandrasekhar's
limit that must go bang in the night or day.) The post-merger product
has the  greatly extended envelope required for a supergiant RCB and HdC star.
 Subsequent 
 evolution follows the canonical post-AGB contraction to the white dwarf track
(\cite{nkr7,nkr8,nkr9}).

In the FF scenario, H-rich post-AGB single stars that
reach the white dwarf cooling track with the potential to experience
a final thermal pulse in their He-shell. Under particular
conditions, the thin surviving H-envelope may experience H-burning.
Onset of the pulse leads to a return of the star to a supergiant-like
form but, if conditions are ripe, this supergiant will be H-poor and
He-rich. Subsequent evolution takes the supergiant across the HR-diagram
to the white dwarf cooling track.
\citet{s1}, \citet{i1}, \citet{re1}, and \citet{he1} among others discuss 
aspects of the FF scenario.  Enigmatic stars such as  FG Sge and 
Sakurai's object (V4334 Sgr) are commonly identified as FF scenario
products.

Quantitative comparison of predictions of the FF and DD
scenario and the observations of the RCB and HdC
stars involves several dimensions. Our focus here is on the
chemical composition obtainable only from examination of the
spectrum of RCB and HdC stars. The (preferably) high-resolution
spectra from the ultraviolet, optical and near-infrared are
analyzed with model atmospheres computed for the
appropriate composition (\cite{nkr2,pandey2006,gar2009}).
Recent elemental abundance studies of C, N, O in particular
seem to favour the DD scenario for the RCBs and HdCs (Asplund
et al. 2000; Saio \& Jeffery 2002; Pandey et al. 2006).   

Two remarkable pieces of  observational evidence
seems to tilt the balance in favour of the DD scenario.
First, \citet{c2005,c2007} made the dramatic discovery that
$^{18}$O is more abundant than $^{16}$O from moderate resolution
spectra of HdC and RCB stars showing the first-overtone CO vibration-rotation
bands at about 2.3 microns. Second, \citet{pandey2006a}  for EHes
and \citet{pandey2008} for RCBs showed that $^{19}$F was drastically
overabundant.  The $^{18}$O and F results show that the DD scenario
was not a quiescent mixing process but either involved
nucleosynthesis during the merger and/or subsequently during the
evolution of the merged product to and from its supergiant phase.

It is amusing to see how this discovery of $^{18}$O was made in HD137613, as
told  us by Tom Geballe. 
He got an email from Geoffrey  Clayton at 5:30 p.m on 2004 November 10:
`Hi Tom. I have this spectrum of the first overtone CO bands in the HdC star,
HD 137613. I took a look at it and the bands were split, 
so I said to myself, oh well there's $^{13}$CO, 
but the extra components aren't at the right wavelengths. 
So
I'm a bit mystified. Take a look at the attached plot and see if you 
have an idea of what this is'. To which Tom replied: ` That is amazing!! 
I happen to have a
book of CO wavelengths that I calculated when I was a grad student in the 70s. 
All dv=1 and dv=2 lines of all isotopic species. 
These bandheads match 12C18O!!
The first three bandheads (2-0, 3-1, 4-2) of $^{12}$C$^{18}$O are at 
2.349, 2.378, and 2.408.
I also looked 12C17O and it doesn't match. 
How could a star have as much $^{18}$O as
$^{16}$O? What kind of star is this? Cheers and congrats, Tom (Geballe 2008).

High resolution spectroscopy is essential for properly estimating the isotope 
ratios $^{16}$O to $^{18}$O, $^{16}$O to $^{17}$O (as well as detection of the 
weak lines of $^{19}$F).  Garc\'{\i}a-Hern\'{a}ndez et al. (2009) used a 
spectral resolution of 50000 in their analysis of the 2.3 micron CO 
bands. They confirmed Clayton et al.'s result of a high $^{18}$O overabundance
for  HdCs and the RCB S Aps obtaining a ratio of  $^{16}$O to $^{18}$O of 
0.5 in HdCs  and the ratio  16 for S Aps (Fig.2). The high resolution 
optical spectra obtained with the Harlan J. Smith telescope at the McDonald 
Observatory and with the  Vainu Bappu Telescope at Kavalur led to the discovery 
that $^{19}$F is enhanced by a factor of several hundreds to a 1000 times solar 
in these stars. In passing we note that Harlan J. Smith and M.K. Vainu
Bappu were great friends.

\begin{figure}
\center
\includegraphics[height=7cm,width=9cm]{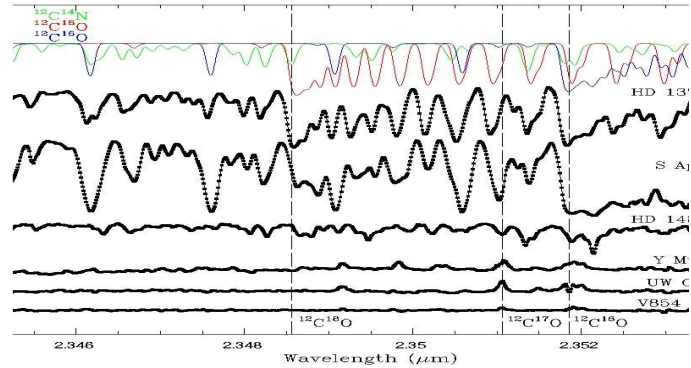}
\caption{High resolution spectra of two HdC stars and four RCBs in 
the region of 2-0 band of $^{12}$C$^{18}$O at 2.3485 microns. 
The top plots show synthetic spectra
computed for the HdC star HD137613 for two
CO isotopologues  and CN. This figure is  from Garc\'{\i}a-Hern\'{a}ndez
 et al. (2009). }
\end{figure}

There can be no doubt that $^{18}$O and $^{19}$F were synthesized in or 
subsequent to the
creation of the RCBs and HdCs. Before full acceptance of the DD scenario
is made, the scenario has to account for the nucleosynthesis of these and
other observable elemental and isotopic abundances.
In principle, as first noted by \citet{warner}, $^{18}$O may be
synthesized by $\alpha$-capture from abundant $^{14}$N. This may occur
during the merger in the DD scenario but not with ease in the
FF scenario.  The reservoir of $^{14}$N is finite and probably
not easily restored, thus, one wonders what cuts off the
high temperatures at the point that there remains an observable
amount of nitrogen.
Other complexities remain. In particular, some RCBs and one HdC show
an appreciable amount of lithium, presumably $^7$Li.
Discussions of the nucleosynthesis in and following the merger are
provided by 
\citet{c2007}, \citet{pandey2008}, and
Garc\'{\i}a-Hern\'{a}ndez et al. (2009). 
Additional high-resolution spectroscopy, especially in the infrared, is
desirable to determine the O isotopic abundances across the RCB sample.

The FF scenario does plausibly account for other stars. Most notable
among the FF candidates are FG Sge, V4334 Sgr
and V605 Aql (Clayton et al. 2005). 
It is not impossible that RCBs from the
FF scenario lurk among the known Galactic, LMC and SMC examples.

\subsection{Dust Production by RCBs}

Historically, RCB stars were the first stars in which circumstellar
dust production was invoked to explain brightness variations, i.e.,
the dramatic declines  of
several magnitudes (\cite{lo,ok}). It is now well established 
that soot (carbon) clouds form close to the star and obstruct the star light
reaching us. Such clouds have now been imaged in the infrared around RY Sgr
at distances of about 1000 stellar radii (\cite{de2004}).  
These clouds were formed presumably closer to the star; observational
constraints prohibited their detection close to the star. 

Questions abound concerning the dust.  What provides the trigger for dust  
production in or around the star?  How may dust condense in or around such 
hot stars as RCBs?  The trigger appears to be related to the atmospheric 
pulsations which are revealed by light and radial velocity variations.
\citet{pu1977} noted a correlation between the beginning of a light decline 
and pulsation phase in RY Sgr.  It has now been shown by \citet{cr2007}
for four stars, by longterm monitoring of the brightness, that when a decline 
happens it does so at a particular phase although not every pulsation
results in a decline.  \citet{Wo1996} developed a model for dust formation 
following   a pulsation-induced shock in the upper atmosphere. When the 
pulsation amplitudes are large, the temperatures and densities in the 
post-shock gas are predicted to be conducive  for  nucleation to occur. 
The question intriguing  spectroscopists  is --   how  may
we  observe a strong shock develop from a pulsation?
What  evidence  may we offer that 
dust condensation actually takes place in the atmosphere of the star?
Can we detect the presence of cool gas from which dust could condense?
High-resolution spectroscopic observations prior to the onset of the 
minimum are crucial to studying such questions.

Recently, R CrB has entered a prolonged light decline that  commenced 
on July 2007 and has not yet ended (July 2009). This is  one of the 
deepest minima the star has had in several decades, 
reaching more than 9 magnitudes. At minimum the star shows a variety of 
emission lines including broad (fwhm of 250 km s$^{-1}$) emission lines of 
He\,{\sc i}, H \& K lines of Ca \,{\sc ii}, Na\,{\sc i} D lines (Fig.1) 
and sharp (fwhm $\le$ 15 km s$^{-1}$) emission lines due to Fe\,{\sc ii}, 
Ti\,{\sc ii} etc. (\cite{Rao1999}). We postpone the discussion of  this 
intriguing emission spectrum for a later occasion.  We were fortunate  to
catch the star just on the verge of the decline.
Our optical spectra obtained at the  McDonald Observatory and at Vainu Bappu
Observatory have a spectral resolution over 60000. Fig. 3 shows the light
curve with the dates  of our   spectroscopic observations marked.
The spectral changes that R CrB  underwent even prior to light 
minimum may provide important clues to the phenomenon of dust production.

\begin{figure}
\center
\includegraphics[height=7cm,width=9cm]{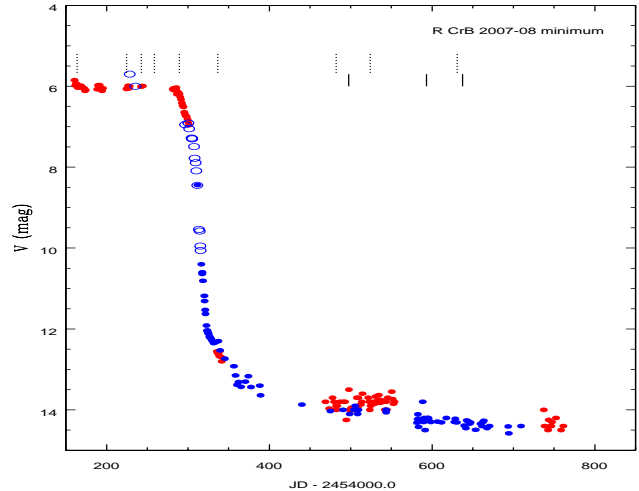}
\caption{
The  visual (red dots) light curve of R CrB from  2007 March to 2008 October. 
Blue dots refer to V magnitudes, blue circles are either no-filter or unknown 
filter measurements, red dots are daily averages of the visual estimates. All 
the data are  obtained from AAVSO data base.  The minimum started around 2007 
July 10.  Dates on which  high-resolution spectra  were obtained are indicated 
by  the dashed lines at the top.  The dates on which low-resolution spectra 
from the HCT are shown as short lines.
}
\end{figure}

\section{ Prelude to Minimum: The Spectrum at Maximum Light}
The present series of spectra that precede the light minimum starts on
2007 March 4. Spectra at maximum light are dominated by lines of C\,{\sc i}, 
O\,{\sc i}, N\,{\sc i},  Si\,{\sc i} and S\,{\sc i}  in addition to the many 
lines of neutral and ionized metals seen in spectra of normal F-G supergiants.
 
The star remained at maximum until 2007 July 10 when the decline in light 
started. In the following account, we discuss briefly the spectral
changes that took place during  maximum light. Even four months prior to the 
onset of the minimum, the star appears to be disturbed with emission in some 
line cores suggesting  a component with an inverse P-cygni profile. Line 
doubling of the absorption lines is  seen two months prior to the onset. The 
blue and red components of the doubled lines showed a different level of 
excitation for the rising gas ($T_{\rm exc}$ of 4300 K) from the infalling 
gas  ($T_{\rm exc}$ of 5600 K).
  
Some lines, like Ti\,{\sc ii} 5154 $\AA$, are not doubled but show emission in 
the line core.  Even the cores of the  Na\, {\sc i} D lines show emission 
compared to the reference spectrum at `normal' maximum light. Moreover, the 
radial velocity of the emissions is same as that of the star. 

The absorption lines  also  broadened without altering the equivalent
widths, a month and half before the onset of minimum as though 
the starlight was passing through a scattering medium (see Fig. 4).
Study  of such line broadening requires high spectral resolution.   

Some aspects of the spectroscopic changes months ahead of the
decline beginning in 2007 July are not unusual. For example, line
doubling appears to be a regular feature of stars such as R CrB and RY Sgr where
the atmospheric pulsation is seen clearly. It will be of interest
to see if these changes occur at the same or similar phases in the
pulsation. Perhaps, there is a  quasi-steady amplification of the changes at the
particular phase lasting a sequence of several pulsations and culminating
in the physical conditions amounting to the trigger for a decline.
Examination of this idea will call for routine observations of a RCB
such as R CrB,  RY Sgr or V854 Cen, a star prone to frequent declines.
Considerable observing time may be wasted in that the declines are
unpredictable and irritation of Telescope Allocation Committees
seems assured. But these datasets are most likely to reveal
new phenomena that may tighten the link between pulsation and
the trigger for dust production. For example, the spectroscopic
variations -- line asymmetry, doubling, radial velocity and strength --
may be more extreme at particular phases. 

More significantly,  major disturbances were present in the spectrum 
of R CrB three  
days prior to the descent.  Comparison of pre-maximum and 2007 July 7 spectra 
shows that most lines in 2007 July 7 spectrum have shifted-absorption 
components (Fig. 5) in addition to the usual photospheric lines. Fig. 5 shows 
a comparison of the spectrum on 2007 July 7 with respect to an undisturbed 
spectrum obtained on 1995 August 17 illustrating the additional absorption 
components to several lines that appear in the 2007 July 7 spectrum.
However, the C\,{\sc i} lines do not show such  additional absorption 
components.  Note that, while the neutral metal lines (e.g.,  Fe\,{\sc i}, 
and Cr\,{\sc i}) show strong blue-shifted components, the ionized metal lines  
(e.g., Fe\,{\sc ii}, and Cr\,{\sc ii}) show red-shifted components on 2007 
July 7.  Such  differences suggests that  rising and falling gases
have different levels of ionization. As further shown in Fig. 6 during the
descent to minimum the blue (ionized) and redshifted (neutral) components
are   separated  by  53 km s$^{-1}$ probably as a result of passage of a
strong shock.  Of seemingly particular relevance to dust production, 
the 2007 July 7 spectrum also displays lines of the C$_{\rm 2}$ Phillips system
showing the presence of gas  at  temperatures (1000 to 800 K, 
Figs. 7 and 8) suggestive of a site for dust nucleation.   
One supposes these unusual line profiles and the appearance of
cool gas are manifestations of the shock induced by the
pulsation that triggered the decline. 

\begin{figure}
\center
\includegraphics[height=7cm,width=9cm]{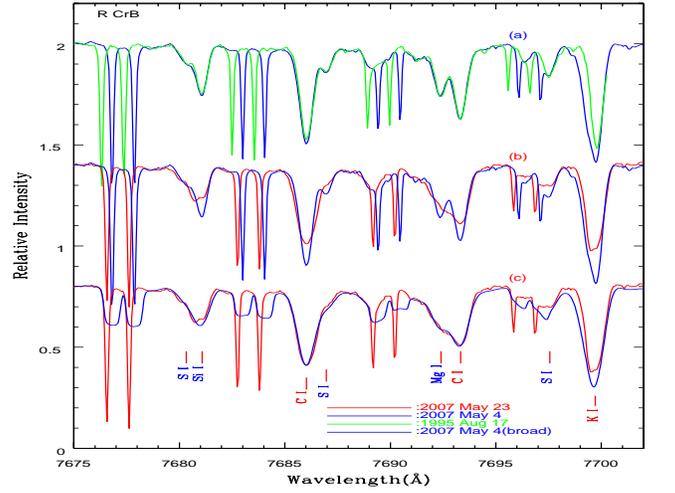}
\caption{
Figure illustrates the broadening of stellar spectrum prior to the onset of the
light minimum. (a) The spectrum on 2007 May 4 (blue) is similar to a typical
maximum one (green - 1995 August 17).
(b) shows the spectral lines on 2007 May 23 (red) are broader 
than in the spectrum obtained 19 days earlier (blue). In (c) the 2007 May 4 
spectrum is artificially brodened without changing the equivalent widths to 
match the 2007 May 23 spectrum.  Only the line shapes got altered.
}
\end{figure}

\subsection{2007 July 7-10 --At the edge of the descent}

\begin{figure}
\center
\includegraphics[height=7cm,width=9cm]{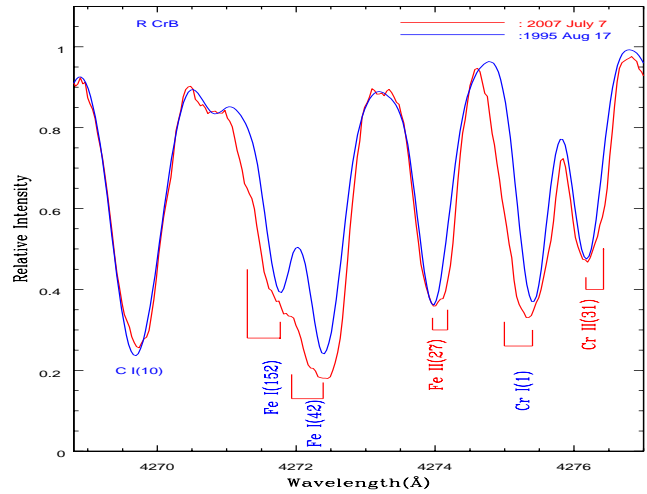}
\caption{
Comparison of the spectral  lines of C\,{\sc i}, Fe\,{\sc i}, Fe\,{\sc ii},
Cr\,{\sc i}, and Cr\,{\sc ii}  in the disturbed spectrum on 2007 July 7(red), 
prior to the descent in light, with a normal undisturbed spectrum on 1995 
August 17 (blue). Note various additional absorption components to both ionized
and neutral lines in 2007 July 7.
}
\end{figure}

\begin{figure}
\center
\includegraphics[height=7cm,width=9cm]{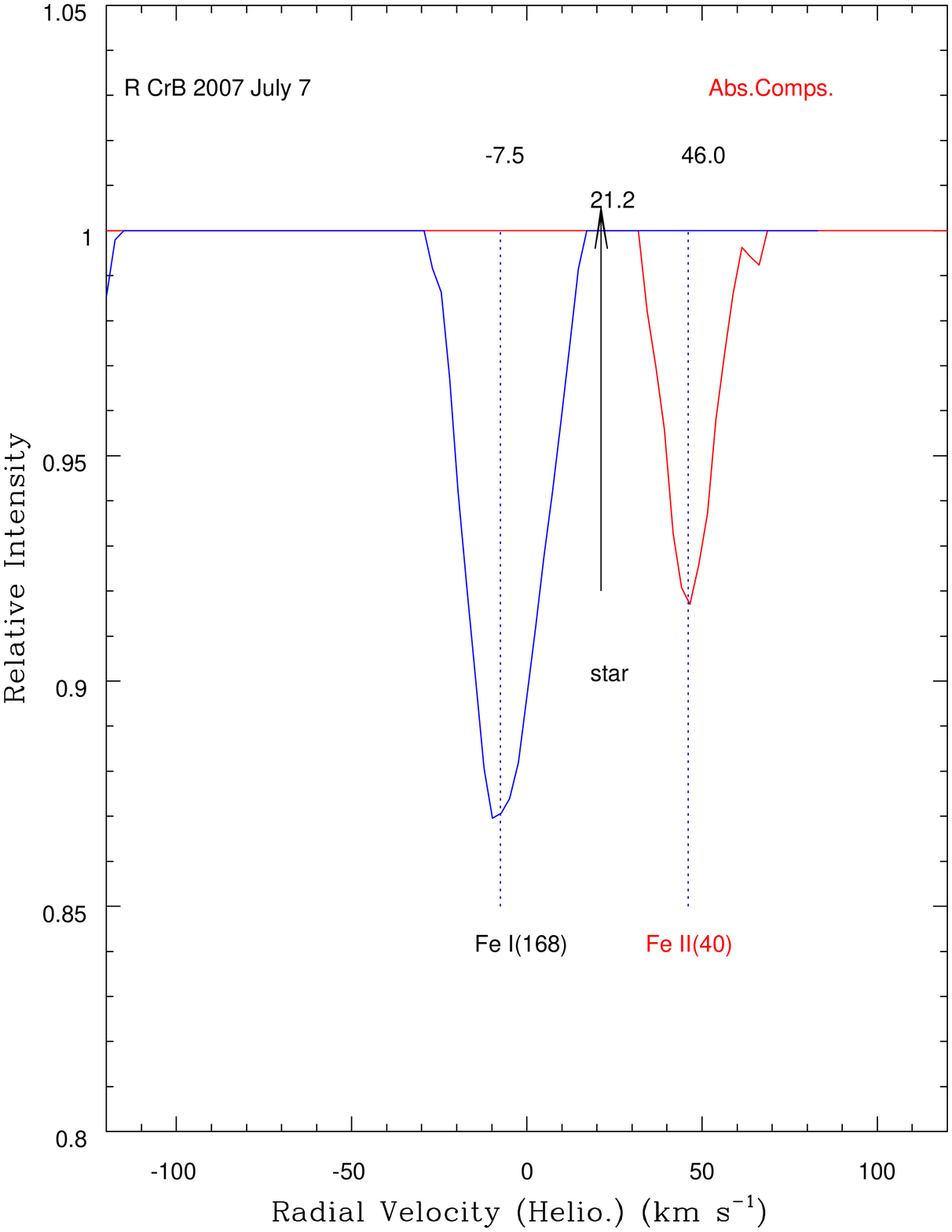}
\caption{
Typical velocities of the extra absorption components to neutral 
(Fe\,{\sc i}-blue-shifted) and ionized (Fe\,{\sc ii}- red-shifted) lines on 
2007 July 7.  The arrow denotes the radial velocity of the star measured from 
high excitation lines without components on 2007 July 7. Note the infalling and
expanding layers have a velocity difference over 50 km s$^{-1}$ that represents 
the atmospheric shock velocity.
}
\end{figure}

\subsubsection{Absorption lines of C$_{\rm 2}$ Phillips system}
The $^{12}$C$_{\rm 2}$ Swan  bands are fairly strong in the spectrum of R CrB.  
Bandheads of the 0-0 band at 5165 \AA, the 1-0 band at 4737 \AA, and the 
0-1 band at 6174 \AA\ are readily identified. More interestingly, the spectrum
of 2007 July 7 shows weak bands of the Phillips system of C$_{\rm 2}$  in 
absorption -- see Fig. 7 for examples of lines from the 2-0 band.  These 
C$_{\rm 2}$ lines  not present in the regular stellar spectrum are shifted in 
velocity from the stellar lines and present with a red and a blue component.
Our spectra  show that the Phillips lines were not present on 2007 June 6 
spectrum, a month before the onset of the decline.

\begin{figure}
\center
\includegraphics[height=7cm,width=9cm]{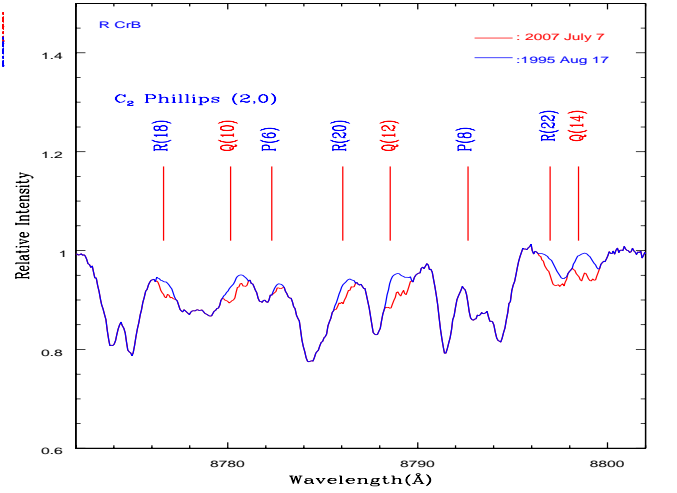}
\caption{
The spectrum of R CrB on 2007 July 7 (red)  in the region 8780 $\AA$ superposed
on the reference spectrum obtained on 1995 August 17 (blue). Absorption lines 
of the $^{12}$C$_{\rm 2}$ Phillips 2-0 band are seen as extra absorptions in 
the 2007 July 7 spectrum but are absent from the 1995 August 17 spectrum.  Line 
doubling in the C$_{\rm 2}$ lines can also be seen.
}
\end{figure}

\begin{figure}
\center
\includegraphics[height=7cm,width=9cm]{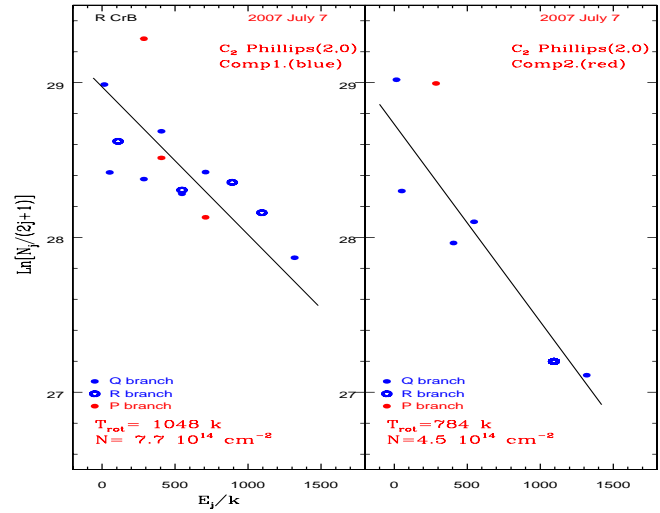}
\caption{
Boltzmann plots of the red and blue  absorption components of C$_{\rm 2}$ 2-0 
lines of the Phillips system from the spectrum of R CrB obtained on 2007 
July 7.  Blue filled circle, open circle and red filled circle symbols refer 
to Q, R and P branch lines, respectively.
}
\end{figure}

Rotational temperatures of 1048$\pm$30 K and 784$\pm$30 K were obtained 
(Fig. 14) for the blue component at  $-$14.5 km s$^{-1}$  and the red 
component at $+$5 km s$^{-1}$,  respectively, using the molecular data given 
by \citet{nkr3}.  The corresponding molecular  column densities are 
7.7$\times$10$^{14}$ cm$^{-2}$ and 4.5$\times$10$^{14}$ cm$^{-2}$.  The radial 
velocities of  C$_2$ molecules of  $-$14.5 km s$^{-1}$ and $+$5 km s$^{-1}$ 
suggest  cool gas rising relative to the star with stellar lines giving the 
radial velocity of 21 km s$^{-1}$.  The cooler gas is less rapidly rising than 
the warmer molecular gas.

These rotational temperatures show that there was gas at the condensation 
temperatures conducive to formation of carbon soot.  Similar detections of cool
C$_2$ molecules have been reported by us for V854 Cen (\cite{Rao2000}),  
R CrB (\cite{Rao2006}) , and V CrA (\cite{Rao2008}). But these detections were 
during  minimum light. In contrast, the present detection of cool C$_2$ 
molecules at maximum light but on the verge of a minimum is significant and 
leads to the conclusion that the light minima are indeed caused by formation of
dust grains. Woitke et al's (1996) model seems to be consistent with 
observations!  Intense spectroscopic coverage of the days leading up to a 
descent to a minimum is an important unfulfilled challenge made more
by difficult by the unpredictable onset of the descent.

\section{The stellar wind}
      
A mark that distinguishes RCBs from HdCs is the absence from the latter's 
defining characteristics, the declines -- deep or even shallow -- that
are such a strong feature of the RCBs. Absence of declines accounts
for the lack of an infrared excess from the HdCs.  Another distinguishing mark 
between RCBs and HdCs is the presence of a stellar wind from RCBs but not from 
HdCs.  Are these two marks different sides of one difference between the
two groups of H-deficient stars?  \citet{ca2003} reported the presence of a 
P-Cygni profile for the He\,{\sc i} 10830 \AA\ line at maximum light for 
several RCBs.  These observations show that the RCBs possess a stellar wind 
with an outflow velocity of 200 to 300 km s$^{-1}$.  Then, Rao, Lambert \& 
Shetrone (2006) showed that in R CrB the strong photospheric lines,
particularly the  O\,{\sc i} 7771 \AA\ line, had a pronounced blue
wing suggesting a component with  an expansion velocity of 120 km s$^{-1}$.
Examination of profiles of other lines showed a component associated
with the stellar wind with the velocity proportional to the excitation energy 
of the lines, i.e.,  the range of lines sample the region in the upper 
atmosphere where the wind is accelerating to the (possibly terminal) velocity 
measured from the He\,{\sc i} 10830 \AA\ line.  Our high-resolution spectra 
now show that a stellar wind may be common, perhaps ubiquitous, among RCBs. 

The temporal dependence of the RCB stellar wind is not yet fully known. There 
is no indication that it depends greatly or at all on the phase of the 
pulsation cycle. Long-term behaviour is also not known.  Observations at the 
1995--1996 minimum of R CrB suggest that the wind-affected wings of the
O \,{\sc i} lines are undisturbed even though the core is affected by 
transient emission in the early decline (Rao et al. 1999). What drives this
wind?  Is there any connection with magnetic fields? In the R CrB wind,
low excitation lines (e.g., the  Al\,{\sc i} resonance lines) show variability 
on a short time scale but there is no change in the high excitation lines.
 Examination of  magnetically-sensitive and magnetically-insensitive
Fe\,{\sc i} lines of comparable strength and excitation potential
from a series of high-resolution spectra of R CrB taken at maximum light
 suggests the variability in the former, but not in the latter (Rao 2008). 
This result suggests a connection between surface magnetic field 
and the wind. The relation between the (apparently) permanent and
(seemingly) pulsation phase-independent wind and trigger for a
decline is as yet unknown.

In contrast, \citet{nkr4} examined the
five known HdCs at the He\,{\sc i} 10830 \AA\ line on high-resolution
spectra  and found no evidence of a stellar wind. The absence of a
wind and the lack of deep declines among the HdCs  and the
presence of a wind among the RCBs with their deep declines is
suggestive that the wind or the presence of an extended atmosphere
as a result of the wind is a necessary condition for occurrence of  
deep declines.

\begin{figure}
\center
\includegraphics[height=7cm,width=9cm]{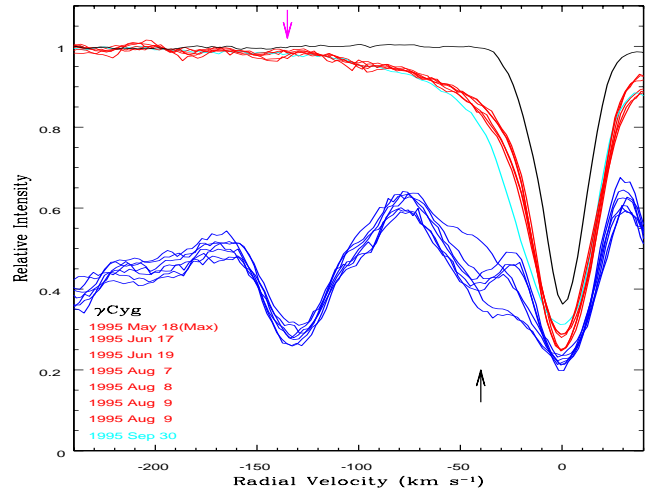}
\caption {
Profiles of O\,{\sc i} 7771 \AA\ (red) and Al\,{\sc i} 3944 \AA\ (blue) in 
R CrB obtained at different times and adjusted to the stellar velocity.  Note the
variability in the low excitation Al\,{\sc i} profiles and the unchanged high 
excitation O\,{\sc i} line profile.  The wind velocity suggested by these lines
50 km s$^{-1}$ for Al and 120 km s$^{-1}$ for O. O\,{\sc i} line profile of
a normal supergiant $\gamma$ Cyg (black) is shown for comaprison.
}
\end{figure}

\section{Concluding remarks}
Answers are slowly emerging for  the important questions posed by RCB and HdC 
stars  concerning the origin of these H-deficient stars, and the mechanisms by 
which a cloud of soot forms and obscures the star. Evidence suggests that the 
DD rather than the FF scenario provides a superior accounting for the elemental
abundances of C, N and O for RCB stars and their likely relatives, the extreme 
Helium (EHe) stars and cool hydrogen deficient (HdC) stars.  The FF scenario 
does plausibly account for other stars. Most notable among the FF candidates 
are FG Sge  and V4334 Sgr, also known as Sakurai's object and V605 Aql 
(\cite{ca2006}).  It is not impossible that RCBs from the FF scenario lurk 
among the analysed sample attributed in the main to the DD scenario.

The process of soot formation is most likely associated with the atmospheric 
pulsation which, on occasions at a certain phase of the pulsation, leads to a 
stronger than usual shock in the atmosphere such that the physical conditions 
are conducive to molecule formation and dust nucleation, as envisaged by Woitke
et al.(1996).  How the dust grains grow  and soot clouds form represent 
questions for future study.

As our essay has hopefully shown, high-resolution optical and infrared spectra 
have and will prove crucial to addressing the leading questions about the 
H-deficient RCB and HdC stars.  In preparing the paper, we came across a 
collection of articles on modern high-resolution spectroscopic techniques.
There, the leading chapter by Sir Harry Kroto carried the title `Old 
spectroscopists forget a lot but they do remember their lines' (\cite{Kr2009}).
This pair of old spectroscopists remember at least the important lines.

\begin{acknowledgement}
We acknowledge with thanks the variable star observations from the AAVSO  
database.  This research has made use of the SIMBAD database, operated
at CDS, Strasbourg, France. Our sincere thanks to the editors of this volume
for accepting with admirable patience our delay in submitting this article. 
This research has been supported in part by a grant (F-634) to DLL from
the Robert A. Welch Foundation of Houston, Texas. 
\end{acknowledgement}

\begin{small}

\end{small}

\end{document}